\documentclass[11pt,twocolumn]{article}
\usepackage{graphicx}
\usepackage{hyperref}
\usepackage{natbib}

\catcode`_=\active
\newcommand_[1]{\ensuremath{\sb{\mathrm{#1}}}}
\usepackage{color}

\newcommand{\advice}[1]{ 
{\begin{quote}
{\color{blue} \bf \em #1}
\end{quote} 
}}
\newcount\longrefs
\def\aap{\ifnum\longrefs=1 {Astron.\ Astrophys.}\else 
                           {A\hbox{\rm \&}A}\fi}
\def\aapr{\ifnum\longrefs=1 {Astron.\ Astrophys.\ Rev.}\else 
                            {A\hbox{\rm \&}AR}\fi}
\def\aaps{\ifnum\longrefs=1 {Astron.\ Astrophys.\ Suppl.}\else 
                            {A\hbox{\rm \&}A Suppl.}\fi}
\def\aj{\ifnum\longrefs=1 {Astron.\ J.}\else 
                          {AJ}\fi} 
\def\ao{\ifnum\longrefs=1 {Applied Optics}\else 
                           {Appl.\ Opt.}\fi} 
\def\aspcs{\ifnum\longrefs=1 {Astron.\ Soc.\ Pacific Conf. Series}\else 
                           {ASP Conf.\ Ser.}\fi} 
\def\apj{\ifnum\longrefs=1 {Astrophys.\ J.}\else 
                           {ApJ}\fi} 
\def\apjl{\ifnum\longrefs=1 {Astrophys.\ J. Lett.}\else 
                            {ApJ}\fi} 
\def\aplett{\ifnum\longrefs=1 {Astrophys.\ J. Lett.}\else 
                            {ApJ}\fi} 
\def\apjs{\ifnum\longrefs=1 {Astrophys.\ J. Suppl.}\else 
                            {ApJS}\fi}
\def\apss{\ifnum\longrefs=1 {Astrophys.\ and Space Science}\else 
                            {Astrophys.\ Space Sci.}\fi}
\def\araa{\ifnum\longrefs=1 {Ann.\ Rev.\ Astron.\ Astrophys.}\else 
                            {ARA\hbox{\rm \&}A}\fi}
\def\azh{\ifnum\longrefs=1 {Astronomicheskii Zhurnal}\else 
                            {Astron.\ Zhur.}\fi}
\def\baas{\ifnum\longrefs=1 {Bull.\ Am.\ Astron.\ Soc.}\else 
                            {BAAS}\fi}
\def\bain{\ifnum\longrefs=1 {Bull.\ Astronom.\ Institutes Netherlands}\else
                            {Bull.\ Astr.\ Inst.\ Neth.}\fi}
\def\gca{\ifnum\longrefs=1 {Geochim.\ Cosmochim.\ Acta}\else 
                           {Geochim.\ Cosmochim.\ Acta}\fi}
\def\grl{\ifnum\longrefs=1 {Geophys.\ Res.\ Lett.}\else 
                           {Geoph.\ Res.\ Lett.}\fi}
\def\iaucirc{\ifnum\longrefs=1 {IAU Circulars}\else 
                          {IAU Circ.}\fi}
\def\ip{\ifnum\longrefs=1 {in press}\else 
                          {in press}\fi}
\def\jgr{\ifnum\longrefs=1 {J.\ Geophys.\ Res.}\else 
                           {J.\ Geophys.\ Res.}\fi}  
\def\jqsrt{\ifnum\longrefs=1 {J.\ Quant.\ Spect.\ Rad.\ Transfer}\else 
                           {JQSRT}\fi}  
\def\jrasc{\ifnum\longrefs=1 {J.\ Royal Astron.\ Soc.\ Canada}\else 
                           {JRAS Can.}\fi}  
\def\mnras{\ifnum\longrefs=1 {Mon.\ Not.\ Roy.\ Astron.\ Soc.}\else 
                             {MNRAS}\fi} 
\def\memsai{\ifnum\longrefs=1 {Mem.\ Ast.\ Soc.\ It.}\else 
                             {MEMSAI}\fi} 
\def\nat{\ifnum\longrefs=1 {Nature}\else 
                           {Nat}\fi}
\def\pasj{\ifnum\longrefs=1 {Pub.\ Astron.\ Soc.\ Japan}\else 
                            {PASJ}\fi} 
\def\pasp{\ifnum\longrefs=1 {Pub.\ Astron.\ Soc.\ Pacific}\else 
                            {PASP}\fi} 
\def\physscr{\ifnum\longrefs=1 {Physica Scripta}\else 
                            {Phys.\ Scrip.}\fi} 
\def\planss{\ifnum\longrefs=1 {Planetary \& Space Science}\else 
                            {Plan. \& Space Sci.}\fi} 
\def\procspie{\ifnum\longrefs=1 {Proc.\ SPIE}\else 
                            {Proc.\ SPIE}\fi} 
\def\qjras{\ifnum\longrefs=1 {Quarterly J.\ Royal Astron.\ Soc.}\else 
                            {QJRAS}\fi} 
\def\sa{\ifnum\longrefs=1 {Soviet Astron..}\else 
                               {Sov.\ Astron.}\fi}
\def\skytel{\ifnum\longrefs=1 {Sky \& Telescope}\else 
                            {Sky \& Tel.}\fi} 
\def\solphys{\ifnum\longrefs=1 {Solar Phys.}\else 
                               {Solar Phys.}\fi}
\def\ssr{\ifnum\longrefs=1 {Space Science Rev.}\else 
                               {Space\ Sci.\ Rev.}\fi}

\def\bibfiles{rjrfiles,adsfiles}   %% RJR + ADSbibfiles on scratch

\def\references{\longrefs=1  \bibliographystyle{rrbib}
             \bibliography{journals,\bibfiles}}

\def\dutch{\def\refname{Referenties}\def\abstractname{Samenvatting}%
  \def\bibname{Bibliografie}\def\chaptername{Hoofdstuk}%
  \def\appendixname{Bijlage}\def\contentsname{Inhoudsopgave}%
  \def\listfigurename{Lijst van figuren}\def\listtablename{Lijst van tabellen}%
  \def\indexname{Index}\def\figurename{Figuur}\def\tablename{Tabel}%
  \def\partname{Deel}\def\enclname{Bijlage(n)}\def\ccname{Ter attentie van}%
  \def\headtoname{Aan}\def\headpagename{Pagina}%
  \def\today{\number\day\space\ifcase\month\or januari\or februari\or maart\or%
     april\or mei\or juni\or juli\or augustus\or september\or oktober\or%
     november\or december\fi \space\number\year}%
  \typeout{
              >>>>> use hlatex209 for Dutch hyphenation <<<<< 
         }}
\newcounter{onefig} \newcounter{fignumber}
\newcount\nocaptions \newcount\nofigures \newcount\figwidth
\newcount\viewgraphs
  \def\paper{}  \def\figlabel{} 
\long\def\nextfig#1{\setcounter{figure}{\value{fignumber}}
  \addtocounter{fignumber}{1}
  \ifnum \viewgraphs=1 \newpage \pagestyle{empty} \fi 
  \ifnum\value{onefig}=0 #1 \fi                 
  \ifnum\value{onefig}=\value{fignumber} #1 \fi}
\def\figwidths#1#2{\ifnum \nocaptions=1 #2mm \else #1mm \fi}  
\def\paper#1{}  %% redefine for separate-figure identification line
\long\def\plotfig#1#2{\ifnum \nofigures=1 \else #2 \fi}
\long\def\captiontext#1{\ifnum \nofigures=1 \raggedright \fi 
   \ifnum \nocaptions=1 \paper
     \ifnum \viewgraphs=0 
       \newline  \mbox{}\hrulefill\mbox{} \newline 
       \newline label:~\{\figlabel\} 
     \fi 
     \else \ifnum \nofigures=0 \fi 
   #1 \fi}

\newcount\panelwidth \newcount\panelheight 
\newcount\bxmin \newcount\bymin \newcount\bxmax \newcount\bymax
\newcount\tbxmin \newcount\tbymin
\newcount\tpanelwidth \newcount\tpanelheight \newcount\tpdif
\def\panelsize #1,#2;{\panelwidth=#1 \panelheight=#2}  
\def\setbb #1,#2;#3,#4;#5,#6;{% UNITS: bp (from ghostview)
  \tbxmin=#1 \tbymin=#2    %% full box (axis titles) lower left corner
  \bxmin=#3 \bymin=#4      %% bare box (ticks only) lower left corner
  \bxmax=#5 \bymax=#6}     %% upper right corner
\def\barepanel #1{%
  \ifnum\panelheight=0 
    \tpdif=\bymax \advance\tpdif by -\bymin
    \multiply \tpdif by \panelwidth
    \tpanelheight=\tpdif
    \tpdif=\bxmax \advance\tpdif by -\bxmin
    \divide \tpanelheight by \tpdif
  \else \tpanelheight=\panelheight \fi
  \psfig{figure=#1,%
     bbllx=\bxmin bp,bblly=\bymin bp,bburx=\bxmax bp,bbury=\bymax bp,clip=t,%
     width=\panelwidth mm,height=\tpanelheight mm}}
\def\labelypanel #1{% TeX permits only integer arithmetic, so bp and mm
  \ifnum\panelheight=0 
    \tpdif=\bymax \advance\tpdif by -\bymin
    \multiply \tpdif by \panelwidth
    \tpanelheight=\tpdif
    \tpdif=\bxmax \advance\tpdif by -\bxmin
    \divide \tpanelheight by \tpdif
  \else \tpanelheight=\panelheight \fi
  \tpdif=\bxmax \advance\tpdif by -\tbxmin
  \tpanelwidth=\panelwidth \multiply \tpanelwidth by \tpdif
  \tpdif=\bxmax \advance\tpdif by -\bxmin
  \divide \tpanelwidth by \tpdif
  \psfig{figure=#1,%
    bbllx=\tbxmin bp,bblly=\bymin bp,bburx=\bxmax bp,bbury=\bymax bp,%
    clip=t,width=\tpanelwidth mm,height=\tpanelheight mm}}
\def\labelxpanel #1{%
  \ifnum\panelheight=0 
    \tpdif=\bymax \advance\tpdif by -\bymin
    \multiply \tpdif by \panelwidth
    \tpanelheight=\tpdif
    \tpdif=\bxmax \advance\tpdif by -\bxmin
    \divide \tpanelheight by \tpdif
  \else \tpanelheight=\panelheight \fi
  \tpdif=\bymax \advance\tpdif by -\tbymin
  \multiply \tpanelheight by \tpdif
  \tpdif=\bymax \advance\tpdif by -\bymin
  \divide \tpanelheight by \tpdif
  \psfig{figure=#1,%
    bbllx=\bxmin bp,bblly=\tbymin bp,bburx=\bxmax bp,bbury=\bymax bp,%
    clip=t,width=\panelwidth mm,height=\tpanelheight mm}}
\def\labelxypanel #1{%
  \ifnum\panelheight=0 
    \tpdif=\bymax \advance\tpdif by -\bymin
    \multiply \tpdif by \panelwidth
    \tpanelheight=\tpdif
    \tpdif=\bxmax \advance\tpdif by -\bxmin
    \divide \tpanelheight by \tpdif
  \else \tpanelheight=\panelheight \fi
  \tpdif=\bxmax \advance\tpdif by -\tbxmin
  \tpanelwidth=\panelwidth \multiply \tpanelwidth by \tpdif
  \tpdif=\bxmax \advance\tpdif by -\bxmin
  \divide \tpanelwidth by \tpdif 
  \tpdif=\bymax \advance\tpdif by -\tbymin 
  \multiply \tpanelheight by \tpdif
  \tpdif=\bymax \advance\tpdif by -\bymin
  \divide \tpanelheight by \tpdif
  \psfig{figure=#1,%
    bbllx=\tbxmin bp,bblly=\tbymin bp,bburx=\bxmax bp,bbury=\bymax bp,%
    clip=t,width=\tpanelwidth mm,height=\tpanelheight mm}}

\def\CC{\par \vspace*{-2ex} \footnotesize \baselineskip=8pt \begin{verbatim}}

\long\def\startignore #1\stopignore{}   %% use \startignore....\stopignore

\def\setlistparams{         
  \topsep=0.7ex                 %% ADAPT: parskip=0: 0.7;  parskip=1: -1.2ex
  \itemsep=0.7ex                %% space between items
  \leftmargini=3ex}             %% dashes at beginning of line 
\setlistparams                  %% recall after type changes 

\newcounter{alistindex}       %% problems: a)  b) etc

\newcounter{romenumnr}

\newlength{\minipagewidth}

\newsavebox{\boxcontent}
\newcommand{\ovalhead}[1]{
  \unitlength=1cm
  \sbox{\boxcontent}{\mbox{~~{#1}~~}}
  \begin{center}
    \ifdim\wd\boxcontent>6ex 
    \ifdim\wd\boxcontent<8cm 
    \begin{picture}(8,3) \thicklines     
      \put(4.0,0.8){\oval(8,1.6)} 
      \put(0.0,0.7){\parbox{8cm}{
         \begin{center} \usebox{\boxcontent} \end{center}}}
    \end{picture}
    \else \ifdim\wd\boxcontent<12cm 
    \begin{picture}(12,3) \thicklines     
        \put(6.0,0.8){\oval(12,1.6)} 
        \put(0.0,0.7){\parbox{12cm}{
           \begin{center} \usebox{\boxcontent} \end{center}}}
    \end{picture}
    \else
    \begin{picture}(16,3) \thicklines     
        \put(8.0,0.8){\oval(16,1.6)} 
        \put(0.0,0.7){\parbox{16cm}{
           \begin{center} \usebox{\boxcontent} \end{center}}}
    \end{picture}
    \fi \fi \fi
  \end{center}} 

\newcounter{headnr}            
\newcounter{subheadnr}[headnr]
\newcounter{subsubheadnr}[subheadnr]
\def\head #1\par{
  \stepcounter{headnr}                          %% sets subheadnr = 0 too 
  \vspace{2ex} \noindent                        %% 2ex = space above, no *
  {\bf \theheadnr~~~~#1}\\[1ex] \noindent}      %% 1ex = space below
\def\subhead #1\par{  
  \stepcounter{subheadnr}
  \vspace{1.3ex} \noindent
  {\bf \theheadnr.\arabic{subheadnr}~~~#1}\\[0.3ex] \noindent}
\def\subsubhead #1\par{
  \stepcounter{subsubheadnr}
  \vspace{1.0ex} \noindent
  {\bf \theheadnr.\arabic{subheadnr}.\arabic{subsubheadnr}~~~#1}\\ \noindent}

\font\dropfont= cmr12 scaled \magstep5
\def\dropcap#1#2{{\noindent
    \setbox0\hbox{\dropfont #1}\setbox1\hbox{#2}\setbox2\hbox{(}%
    \count0=\ht0\advance\count0 by\dp0\count1\baselineskip
    \advance\count0 by-\ht1\advance\count0by\ht2
    \dimen1=.5ex\advance\count0by\dimen1\divide\count0 by\count1
    \advance\count0 by1\dimen0\wd0
    \advance\dimen0 by.25em\dimen1=\ht0\advance\dimen1 by-\ht1
    \global\hangindent\dimen0\global\hangafter-\count0
    \hskip-\dimen0\setbox0\hbox to\dimen0{\raise-\dimen1\box0\hss}%
    \dp0=0in\ht0=0in\box0}#2}

\def\level #1 #2#3#4{$#1 \: ^{#2} \mbox{#3} ^{#4}$}   

\def\mathstacksym#1#2#3#4#5{\def#1{\mathrel{\hbox to 0pt{\lower 
    #5\hbox{#3}\hss} \raise #4\hbox{#2}}}}

\mathstacksym\lta{$<$}{$\sim$}{1.5pt}{3.5pt} % less than approximately
\mathstacksym\gta{$>$}{$\sim$}{1.5pt}{3.5pt} % greater than approximately
\mathstacksym\lrarrow{$\leftarrow$}{$\rightarrow$}{2pt}{1pt} % equilibrium
\mathstacksym\lessgreat{$>$}{$<$}{3pt}{3pt} %% less or greater

\def\bibfiles{/Users/judge/tex/bibfiles/books,%
              /Users/judge/tex/bibfiles/gfatom,%
              /Users/judge/tex/bibfiles/lineform,%
              /Users/judge/tex/bibfiles/prdpola,%
              /Users/judge/tex/bibfiles/caiimgii,%         
              /Users/judge/tex/bibfiles/modcont,%
              /Users/judge/tex/bibfiles/rotwaves,%         
              /Users/judge/tex/bibfiles/finstruc,%
              /Users/judge/tex/bibfiles/instrum,%          
              /Users/judge/tex/bibfiles/general,%          
              /Users/judge/tex/bibfiles/rjr,%
              /Users/judge/tex/bibfiles/allref,%          
              /Users/judge/tex/bibfiles/atoms,%          
              /Users/judge/tex/bibfiles/mc,%          
              /Users/judge/tex/bibfiles/spectra,%          
              /Users/judge/tex/bibfiles/zeeman,%          
              /Users/judge/tex/bibfiles/judge}

\def\references{\bibliographystyle{/Users/judge/tex/styles/rrbib}
                \bibliography{/Users/judge/tex/bibfiles/journals,\bibfiles}}

\begin{document}

%###############################################################################
%
%     OPENING
%
%###############################################################################

\title{On choosing meaningful research projects in the natural 
sciences\footnote{This article was edited and condensed for publication in {\em Physics World}, and entitled ``Starting out strong'':
\url{http://iopscience.iop.org/article/10.1088/2058-7058/30/10/36/meta}
}
}
\author{Philip G. Judge, Isabel Lipartito and Roberto Casini}

\maketitle

%###############################################################################
%
%     ABSTRACT
%
%###############################################################################

\begin{abstract}
Over the last few decades,
the nature of scientific research has changed
in response to external influences.  Firstly, powerful
networked computers have become a standard tool.  Secondly, society presses ever harder for
research to deliver something
``useful" back to society, both through the
kinds of funding opportunities that are made available, and through a critical public eye.  Many funding 
agencies now demand ``deliverables'' that seem
to select research of a particular kind.   Lastly,
teamwork, often within very large projects, has become commonplace. 

Here, we step back and ask how prospective research scientists might select productive research projects in this evolving environment. We hope that our suggestions might also help to improve public understanding and thereby restore flagging faith in science.

\end{abstract}

\section{Motivations}

%Can and should natural science  survive as an %independent and worthwhile endeavor?
Researchers in the natural
sciences usually feel privileged.  Excited about our
subjects, thrilled by the prospect of stepping into the unknown, we count ourselves lucky
when we can support our families by doing something we enjoy --  research.  Motivated by our
belief that research in natural science is a noble and productive pursuit, something
worth preserving for future generations, this 
informal article aims to help those considering science careers to decide if research is for them, and if so, what kind of research. 

We are also motivated to try to understand what it is that makes a good
research project.  Increasing numbers of young researchers seem to struggle to answer the
 question, ``why should one care about your
research?''  The question is not meant unpleasantly, it is asked to provoke thought about the motivations behind the research project in relation to the more abstract 
desire to advance knowledge of nature.

%This has lead us to reflect on the nature and %motivation behind modern research projects. 
%We begin by looking at a little history of the %relationship between scientific advancement %and institutional or public support.

\section{Science, technology, research}

We must first specify our terminology. In modern culture, {\em science} and {\em technology} go
hand-in-hand.  For example, NASA is known, rightly, as both a
scientific and technological institution.  High impact publications such as {\em Science} and
{\em Nature} report both scientific and technological advances.  But what do we mean by {\em research} in technology and {\em research} in science?

Most people might agree that research in technology is the pursuit of advancing tools.  The goal of research in technology is improvement of technology itself, not a better understanding of nature. Technological research is not scientific research {\em per se},  because the essence of scientific research is to {\em perform experiments to challenge our understanding of nature}.  

The necessity of experimental arbitration in science is traditionally associated with Galileo Galilei.  Four centuries later, Einstein became famous for ``thought experiments'', mental scenarios rooted in simplicity and symmetry, again meant to discriminate between acceptable and unacceptable theories of how nature works.
Accepting both real and thought experiments as the arbiters for new ideas gives us a working definition of modern research into natural science. 

Karl Popper 
\citep[e.g.][]{Popper1972} suggested further that scientific research must also be {\em falsifiable} or {\em refutable} by experiment. Any advancement must also determine its new domain of applicability which must also be open to refutation through experimentation.  
%Let us accept for the moment Popper's %requirement.  
Not everyone agrees with Popper's viewpoint. But if one embarks on research where falsification seems remote, one 
should be prepared to defend the work in other ways.  Generations of string theorists may have to live with this reality, unless practically testable
predictions can be uncovered.  This possibility is, to the authors' knowledge, non-zero. Research on any theory that might unify physics surely remains of prime importance to natural science, no matter if it can currently be refuted.

Research into natural science has a unique goal.  There is nothing to be ``sold" or ``delivered" except a better understanding of nature.  Curiosity about the interaction of atoms with light led to the discovery of the laser,  the interaction of magnetized atomic nuclei with low frequency radiation led to MRI machines. Neither development could have taken place without the research into the natural science beforehand.  Therefore \advice{curiosity-driven research without foreseeable outcomes remains important.} 
Most modern research will fall between  curiosity-driven and what we will call deliverable-driven research.  
Prospective students should be able to judge where in the spectrum a given research project may lie.  Those focusing on tangible deliverables  necessarily limit the scope of the  research, for \advice{if one already knows the outcome (i.e., promises deliverables), the research must usually be of a technical, not natural scientific nature.} 

These points are not meant to be judgmental, after all, these are all human activities and they are necessarily imperfect. But prospective scientists might ask themselves:  
\advice{Am I really interested in science or technology, or both?}  
To some, an answer might help in making an important decision, to others it might not matter.  Indeed many great scientists did or do both (for example, Sadi Carnot, Michael Faraday, Kristian Birkeland), but they are rarer nowadays owing to specialization. Such people  gain much respect from their peers, or they deserve to!

We turn to the meaning of {\em research}, although
the reader will have a good idea of what it entails.  It is worth a look at some definitions. The Oxford English Dictionary states:

\begin{quote} {\em The systematic investigation into and study of materials
and sources in order to establish facts and reach new conclusions.}
\end{quote}
\noindent
From Webster's Dictionary, we have
\begin{quote}{\em 
1. Diligent inquiry or examination in seeking facts or principles;
laborious or continued search after truth... -- Macaulay.\\
2. Systematic observation of phenomena for the purpose of learning new
facts or testing the application of theories to known facts; - also
called scientific research.}
\end{quote}

Implicit in Webster's point~1 is that something is research if it is {\em new}, new to an
individual or to society.  It is in this second sense and in point 2 that 
 {\em professional} research, the subject of this article, must be limited. 

But what makes {\em scientific research?} It would help if there were a
prescription, an established ``scientific method" identified and acknowledged universally.  But that taught in schools is a myth generally dispelled before graduate school.  In reality, there is no universally recognized method
\citep[e.g.][]{Newton-Smith1981}. It can vary from ``anything goes" through ``trial and
error'' to strict hypothesis testing.

History again is useful to consider. 
One view 
presented by Thomas Kuhn in his ``The structure of
scientific revolutions" \citep{Kuhn1970} argues that advances tend to move in a ``science as usual" fashion with
modest changes to existing ideas (this is often dubbed
\emph{incremental science}), until a discovery is made which,
eventually, will overthrow a previous paradigm (quantum mechanics rejecting   determinism, for example).

We believe that prospective scientific researchers owe it to
themselves to try to do research that might just change paradigms.  This
is a controversial issue, since few research organizations can afford to fund, and usually they do not fund,
research that plainly states this as a goal.  For their ``stakeholders", this is just too risky.

How then is a young researcher to decide on their first research
project?  Are they simply to accept the words of wisdom of a potential
advisor?  Most do precisely that!  But make no mistake, your
first or second research project will usually determine your future
career which will hopefully last a lifetime of productive research in science.
So there is a lot at stake.

\section{Risk}

To discover something {\em new}, risk is necessary.  How is one to tell if a research project has a genuine element of risk and potential for discovery?  
One might simply ask:
\advice{What in the project is genuinely  new?}  Any good scientific advisor should happily answer this obvious question without feeling offended. But 
we believe that there there is also a less obvious form of non-newness that has evolved in response to external forces. 
Scientific research is sometimes so well planned as to admit little room for real discovery.  Some research groups thrive by repeating an established process, but with a new twist.  Under pressure from outside to deliver, increasingly "risk-averse" funding agencies often welcome proposals along these lines. Funding agencies also gain credit from development of new tools.  Thus there is a trend towards ``tool driven research'', in which the choice of a research project is driven entirely by new or  older facilities and associated funding opportunities.  Risk-averse work can also be easier to publish, genuinely new ideas often receiving greater scrutiny by referees.   Members of such groups graduate ``on time", with several publications to their name, and a track record of laboratory success, measured by these and other such ``metrics''.  But were originality and innovation inadvertently filtered away by a system that has become risk-averse?  
Therefore we  encourage  prospective researchers to ask the additional question, 
\advice{Does the research allow for truly unexpected outcomes?}
Good advisors will be delighted to hear this question from a potential student and will energetically discuss exciting possible new discoveries.  In other words, we warn  students about advisors who seem to treat graduate school as a revolving door, focused on quantity of results, publications, and other such ``metrics".  For what kind of ``metrics" could one apply to Einstein's research, for example? 

\section{Experimentation}
 Computers are now far more widely used in scientific research 
than any other tool, with the exception of the human 
brain.  
Even the notion of an ``experiment''  is undergoing revision to include  ``numerical experiments'', in which  
 ``data'' (Latin for ``things that are given'')
are generated by a computer program.  This
begs the question, if falsification through experimentation is a
main-stay of modern scientific research, then what is the role of such 
``numerical data''? This issue is a tricky one that prospective researchers in natural science should be aware
of, because of other recent 
trends:
\begin{itemize}
\item  Many researchers are given a numerical model -- a ``code'' -- 
 as a primary research tool. 
\item Others are given data from a remote machine or instrument wholly developed by others with data delivered by a computer.
\item It is increasingly common to pit numerical data directly against instrumental data, without necessarily recognizing associated practical and philosophical problems. 
\end{itemize}
The last point can be straightforward or tricky. Consider the claim that a non-trivial numerical calculation agrees with experiment. Nothing is refuted, the theory survives another day. But several important questions arise. What is the "information content" of the data, in other words, how strong is the connection between the essence of the model and the data?   Could data have been acquired with a stronger connection to the model? Are there other models compatible with the same data?  Have the authors inadvertently ``cherry picked" from computed or experimental data, or both?  Are there areas of disagreement, however minor, and if so, how is this  handled? More generally, how are we to compare data from a virtual calculation with data from nature?

The common use of ``numerical experiments" is relatively recent, and 
the current situation marks a  departure from what 
was considered as standard as recently as three or four decades ago.   Fascinating new
phenomena have been discovered of direct relevance to science and
nature, for example, in the field of non-linear systems. Entirely new subjects including complexity and chaos have emerged, sparked by the  pioneering work of Fermi, Pasta \& Ulam
 (1955, see \citealp{Dauxois2008}).  
These new research areas are directly relevant to natural science.  But 
there remain problems that are
inaccessible to computers and that will remain so for decades to come. A well-known example is the coupling across enormous spatial and
temporal scales needed to address critical problems in weather forecasting, 
atmospheric science, biology, astrophysics, and many other fields.  

We must return to our goal here to offer advice,
not to delve into these difficult but important issues.  Indeed we might risk the wrath of numerical experimenters merely by
pointing these issues, so much of modern research is done using such tools.  But numerical data are {\em not} on the same
footing as actual data, if only for the simple reason that current
computers always return a deterministic solution to a given problem.
There is no genuinely``random'' number generator for a computer, yet nature does just  
this.  Thus a computer can in principle never simulate
reality.  Instead, researchers  mimic randomness
using techniques such as ensembles, multiple
realizations of dynamical systems started from different initial states.

In a more mundane but no less  worrying development, 
some new researchers have to treat a 
computer code as a ``black box'' that is assumed to represent something in nature
itself.  It is as if the computer has become a real, actual
``experiment''.  We are 
not qualified to assess further if computer data can be treated 
as if they were data from nature, so we restrict our advice to the following.  We suggest
that a prospective numerical scientist ask
\advice{what is incomplete and/or missing from
  the numerical model, and how will the calculations  connect 
with reality?}  If no satisfactory answer
is forthcoming, then the prospective numerical scientist
will shoulder much of the burden of making genuine  breakthroughs, and of convincing the scientific community how their work connects with reality. 

We have not been entirely fair in singling out ``numerical
experimentation''.
If a computer and/or code steps into regimes demonstrably new
owing to a technological breakthrough, the chances of doing excellent
and meaningful research is significant.
We are also limited by our brief human lifetimes -- we do not live long enough to witness a galaxy merger, or the full life of a star.  In such cases, analytic theory and numerical calculations can bring us closer to understanding phenomena than we ever could have hoped for otherwise. Perhaps surprisingly, ``real" experimental data also suffer from problems in common with
numerical data.  Every measurement requires a finite time to make, we only ever measure ``averages over time''.  The averages are also usually over space, or they radically under-sample
the space of interest.  Most stars with 
their associated complex phenomena are completely unresolved!  
In the same way that a
numerical code for the dynamics of fluids is limited by 
the grids upon which they are based (or some equivalent parameters),
observations are limited by their resolution or sampling rates in
time, space, wavelength, among other parameters.  

It must also be remembered that
Nature does not always permit us to observe what we, as a society,
might really like to know.  Generally speaking, experimental data carry with them a
certain -- limited -- ``information content'' noted above. We measure photons, electrons, other
particles, bubbles traced by particles. In the area of solar physics, for example,
we need to know how magnetic free energy is stored and released in solar plasma
to produce damaging solar flares.  But this free energy is not
observable, instead we observe integrated signatures that are related 
weakly to it.  

Fortunately, these potential worries, of importance philosophically,
are often pragmatically allayed by asking again the
deceptively simple questions: \advice{What is new,  what
  is the potential for genuine discovery?}  The prospective researcher should
be prepared for quite complex answers, for much work in science is also an art.

\section{Scientific freedom}

Seen from the outside, scientific research can appear to be an ongoing
advancement, planned years before, perhaps like the slow development
of a new town.  But many exciting discoveries are made while
making all kinds of mistakes and blunders.  The mistakes often offer increased understanding.  A good research environment will implicitly give permission to {\em fail}, at least some of the time.  This is perhaps the most freeing part of any job.  So we offer up the following suggestion:
\advice{Give yourself permission to make mistakes and even to fail.}

\begin{figure*}[h]
\begin{center}
\includegraphics[width=0.8\textwidth]{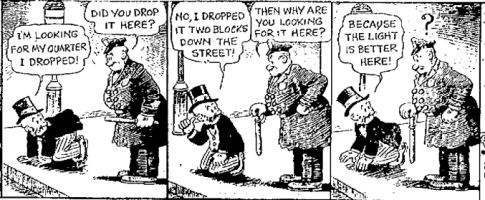}
\end{center}
\caption{Once you have identified the right question, don't be afraid to use or develop the right tools for the problem at hand. Don't be like the old gentleman!}
\end{figure*}

By doing so, you will be able to take on risk.  Research without risk
is not in fact research, because  something ``new" must be discovered which, by definition, is unknown!  Many historical instances 
testify that {\em mistakes and failures are an important part of
  scientific discovery.} The discovery of penicillin by
Fleming is an obvious example.  A case can be made that
overly-planned, limited-risk research can lead primarily to incremental scientific discoveries.  

There is an old cartoon (see the accompanying figure) showing a gentleman looking for a quarter under a street light. By using the tools he knows (the street light), he will never discover something important (the quarter) unless he
takes a risk and tries something new (use a flashlight or metal detector, perhaps).  The gentleman may find something, but he is ``risk-averse", he chooses to use only the tools at hand. Science advances when we advance through the dark, not when we remain under the streetlight.

As noted, funding agencies often 
demand to {\em know} the likely outcome of the research even before
it has been done!  ``We ask to use this nice radio telescope to find
something no-one else has found'' is a proposition unlikely to succeed
during any time in the modern era of astrophysics.  But such ``fishing
expeditions'' are an important part of discovery (see below).  As a
result, many researchers have learned to play a game, that of guessing
outcomes, in order to succeed in gaining funding.  Some actually know
the outcomes before proposing, although no-one will admit this,
because the research has actually been done.  But risk is perhaps the
{\em most important ingredient} needed to do truly meaningful
research.

\section{Research in large projects}

A glance at the numbers of authors on publications reveals another
interesting development.  The days in which an individual scientist
can make significant advances by working alone are getting rarer.
It is easy to think of many examples of huge projects (LHC, Hubble
Space Telescope, the Human Genome Project, ITER,...), as well as many
smaller projects involving many individuals.   
In this state of affairs,  prospective scientists should try to get a
 clear statement of  \advice{how their original work will fit into a large project, and how the advisor plans to protect the student's 
interests.}  Good advisors of course understand this very clearly,
they get funding promising one thing and then, with a very open mind,
perhaps discover something truly new.   So, in addition, \advice{try to find
advisors who welcome latitude in research,} even if they are part of
a much larger project.  

\section{Some final thoughts}

\advice{Get some experience as a summer intern. }  There are many opportunities!  
This will enable prospective researchers to get a feel for the
research environment, and to think on the issue of science vs. technology, or both.

\advice{Choose a research area that you find compelling.}  This may seem obvious, but
there are many bright people who did not succeed because
they simply did not have that dogged determination to solve problems
that really interested them.  Tenacity is a great virtue in research.   For most good researchers,
their research is never simply a ``job". 
\advice{
Find an advisor who does not insist on being on every paper that their
students publish.}    A young scientist who has a single author paper is a rarity
these days.  But these are the only publications where the real mettle of
the person can be judged by an outsider, unless the person is well-known in the community
through presentations, meetings, contacts.  Imagine reviewing yourself for
a tenure track job in say 5-10 years time.  How will you stack up if
you have twenty-odd 
papers as a subordinate author, compared with a few first or a couple of
single author papers?  
\advice{Find an advisor who loves their subject.}   The enthusiasm will be shared
and will make the journey easier.
\advice{Find an advisor who at least 
knows about ``Bayesian" methods.} 
This might sound strange in an article of general interest.  But when
comparing hypotheses (usually encaptured these days in ``models" on a
computer) with experimental data, this question will reveal the seriousness of
the issue of comparison of theory with experiment.  This question lies at the heart of methodologies
in natural science.  Bayesian techniques can help one avoid doing
research of the kind that is ``not even wrong", by forcing you to make some kind of hypothesis and
assess quantitatively how well a given set of observations are
compatible with it.  Bayesian methods are in fact not used as much as
one would think, neither are they necessary. The alternate statistical
description (taught first in colleges) --  ``frequentist" --  draws a conclusion based upon the frequency of the results that lead to this certain conclusion.  But
Bayesian methods {\em require} one to make an hypothesis, up front, to
be tested.  In this sense they automatically satisfy and quantify the
``falsifiability'' test advocated by Popper.

Bayesian methods contrast with 
another perfectly valid kind of scientific research 
unkindly called a ``fishing expedition", for obvious reasons.  
Because there is no single accepted scientific methodology, ``fishing expeditions" are sometimes exactly what is needed.  Of many
successful expeditions throughout history,  Darwin's voyage on 
the Beagle is a good example.  This falls into the ``anything goes" philosophy of science.   Provided that {\em some} earlier barrier has been removed, these studies are essential.

In most subjects these barriers move slowly but surely.  In solar physics, the last ``big breakthrough" was arguably around 1989 when the internal rotation of the Sun was first brought to
light using the technique called helioseismology.  
Since then, we have seen newer, better instruments, gradually stretching out those barriers in
resolution, time-span, energy, measurement precision, something that may
ultimately lead to a new and genuine break-through.  But for now, we are
in more of a phase of ``business as usual" than ``paradigm changing'' 
solar research.  We hope to be astonished by the findings of
young people in the future, in which classes of models might be rejected.
This ``rejection'' automatically satisfies Pauli's desire to do research 
that is definitely not, ``not even wrong''. 

\references

\vskip 0.2truein

\noindent {\bf Acknowledgments.}
We are grateful to Erica Lastufka, 
for helpful comments on the manuscript. 

\vskip 0.2truein

\noindent {\bf About the authors.}

PGJ and RC are senior researchers in physics at NCAR. IL is a PhD student, currently researching new instrumentation for 
astrophysical applications at UCSB. 

\end{document}